\documentclass[aps,pra,twocolumn,floatfix,twoside]{revtex4}

\usepackage{graphicx}
\usepackage{dcolumn}
\usepackage{bm}

\begin{document}

\title{Thermodynamic properties and phase transition of strongly interacting
two-component Fermi gases}
\author{Hongwei Xiong}
\affiliation{State Key Laboratory of Magnetic Resonance and Atomic
and Molecular Physics, Wuhan Institute of Physics and Mathematics,
Chinese Academy of Sciences, Wuhan 430071, P. R. China}
\affiliation{Center for Cold Atom Physics, Chinese Academy of
Sciences, Wuhan 430071, P. R. China} \affiliation{Graduate School
of the Chinese Academy of Sciences}
\author{Shujuan Liu}
\affiliation{State Key Laboratory of Magnetic Resonance and Atomic
and Molecular Physics, Wuhan Institute of Physics and Mathematics,
Chinese Academy of Sciences, Wuhan 430071, P. R. China}
\affiliation{Center for Cold Atom Physics, Chinese Academy of
Sciences, Wuhan 430071, P. R. China}
\author{Mingsheng Zhan}
\affiliation{State Key Laboratory of Magnetic Resonance and Atomic
and Molecular Physics, Wuhan Institute of Physics and Mathematics,
Chinese Academy of Sciences, Wuhan 430071, P. R. China}
\affiliation{Center for Cold Atom Physics, Chinese Academy of
Sciences, Wuhan 430071, P. R. China}
\date{\today }

\begin{abstract}
The thermodynamic properties of two-component Fermi gases with divergent
scattering length is investigated and the transition temperature for the
emergence of a stable dimeric gas is obtained by a simple theoretical model
where the unique property of the dimers (\textit{i.e.} fermionic atom pairs)
with divergent scattering length is considered. Below the transition
temperature, through the investigation of the overall entropy for the
mixture gas of fermionic atoms and dimers, we calculate the relation between
the energy and temperature of the system. In the limit of zero temperature,
based on the chemical equilibrium condition, the fraction of the dimers is
investigated and the role of the dimers in the collective excitations of the
system is also discussed. It is found that our theoretical results agree
with the condensate fraction, collective excitations, transition temperature
and heat capacity investigated by the recent experiments about the strongly
interacting two-component Fermi gases.

PACS numbers: 03.75.Ss, 05.30.Fk, 03.75.Hh, 05.70.Fh
\end{abstract}

\maketitle

\section{Introduction}

It is a quite challenging theoretical problem to explain the basic
properties of the strongly interacting many-body system because there is no
exact solution for this nonlinear many-body system and the ordinary
perturbation method can not give us a reliable quantitative theoretical
predication. The strongly interacting Fermi gases is a very important
theoretical problem because the atomic nucleus, quark-gluon plasma and high
temperature superconductivity \textit{etc} relate closely to the strongly
interacting Fermi gases. Due to the strong interaction, a phenomenological
theory would contribute largely to our understanding of this complex system.
For example, for the atomic nucleus which is a complex many-body system
bound by the strong interaction, there is a successful phenomenological
theory by introducing an effective force. Obviously, clear experimental
investigations can give us important clues to find a reasonable
phenomenological model and even a theory to reveal the microscopic mechanism
accounting for the basic phenomena of the strongly interacting many-body
system. In the last few years, the experimental advances on the ultracold
strongly interacting two-component Fermi gases such as \cite{THOMAS-Fermi}
provide us an important opportunity to develop a theory of the strongly
interacting system.

There is a broad interest in the ultracold two-component Fermi gases because
the two-component Fermi gases can be cooled to a temperature far below the
Fermi temperature so that there is an obvious quantum statistical effect.
Another special interest in this system lies in that through a
magnetic-field Feshbach resonance, one can change in a controllable way the
value and even sign of the scattering length $a$ between atoms. For
two-component Fermi gases, when the magnetic field is tuned so that the
energy of a quasibound molecular state in a closed channel matches the total
energy in an open channel, there is a magnetic-field Feshbach resonance.
Below the Feshbach resonant magnetic field (BEC side), the interaction
between atoms is repulsive and there exists molecule which is a short-range
fermionic atom pairs. Below the critical temperature, there is a molecular
Bose-Einstein condensation (BEC) which has been observed in the recent
experiments \cite{JOCHIM,JIN,MIT-mole}. Above the Feshbach resonant magnetic
field (BCS side), the interaction between atoms is attractive and there
would be a Bardeen-Cooper-Schrieffer (BCS) superfluid behavior due to the
atomic Cooper pairs at sufficient low temperature. Recently, by tuning the
uniform magnetic field near the resonant magnetic field, there are a lot of
interesting experimental works on the BCS-BEC crossover \cite%
{GRIMM,JIN-fermion,MIT-fermion,THOMAS,GRIMM1,Cheng,SALOMON,HEAT} such as the
measurement of the cloud size \cite{GRIMM}, condensate fraction \cite%
{JIN-fermion,MIT-fermion}, collective excitation \cite{THOMAS,GRIMM1},
pairing gap \cite{Cheng}, and heat capacity \cite{HEAT} \textit{etc}.
Besides the pioneering theoretical works \cite{EAGLE,LEGGETT,NOZI}, in the
last few years, there are intensive theoretical investigations on the
BCS-BEC crossover \cite%
{STOOF,TIMM,MILSTEIN,OHASHI1,STAJ,WU,CARR,BRUUN,FALCO,STRI,XIONGCROSS,XIONG-COE, HEISELBERG,CARLSON,HO1,HO,XIONGON,TORMA,LEVIN,CHENG-mean}
such as resonance superfluid \cite{TIMM,MILSTEIN} and universal behavior for
the gases with divergent scattering length \cite%
{HEISELBERG,CARLSON,HO1,HO,XIONGON}.

In the BCS-BEC crossover regime, the scattering length becomes divergent on
magnetic-field Feshbach resonance, thus the system on resonance gives us an
ideal system to investigate directly the strongly interacting many-body
system. In the limit of this strong interaction, the scattering length
becomes divergent and thus it should not appear in the final expression of a
physical quantity. For the ultracold gases with divergent scattering length,
it is a quite attractive theoretical problem considering the possibility
that a phenomenological theoretical model would be very simple because the
scattering length will not appear in the final expression of the physical
quantity although it is the most important parameter to show the interaction
between particles. In the present work, we investigate theoretically the
thermodynamic properties of the two-component Fermi gases in the case of the
scattering length being much larger than the average distance $\overline{l}$
between particles. On the side of weakly attractive (or repulsive)
interaction, there are pairs of atoms in the form of atomic Cooper pairs (or
molecules) which has been verified through the experimental investigation of
the condensation fraction \cite{JIN-fermion,MIT-fermion} and pairing gap
\cite{Cheng}. Generalizing this observation, we assume here that there are
dimers which comprise two fermionic atoms with different spin state for the
system on magnetic-field Feshbach resonance. Most recently, there is a
measurement of the heat capacity \cite{HEAT} for strongly interacting Fermi
gases of $^{6}Li$ atoms which gives clearly the evidence of a phase
transition for the emergence of the fermion pairs. In the present work, at
finite temperature as well as zero temperature, the thermodynamic properties
of the system such as the entropy, transition temperature, fraction of
dimers are investigated based on a simple theoretical model that the system
is in a mixture of the Fermi gases and dimeric gas below the transition
temperature.

In Sec. II, at zero temperature, we consider the thermodynamic properties
for the mixture gases of fermionic atoms and dimers based on the chemical
equilibrium condition. It is found that this theoretical model is consistent
with the recent experimental investigation of the condensate fraction \cite%
{MIT-fermion} and collective excitations \cite{THOMAS,GRIMM1}. The
coexistence of Fermi gases and dimeric gas at zero temperature means that
there should be a phase transition for the emergence of dimeric gas at
finite temperature. In Sec. III, when the thermal excitation of the dimers
is considered, we give a transition temperature that below this transition
temperature there would be a stable dimeric gas. The fraction of dimers
below the transition temperature is also given. In Sec. IV, through the
investigation of the overall entropy of the Fermi gases and dimeric gas, we
calculate the relation between the energy and temperature of the system. The
role of the dimeric gas is discussed and found that the theoretical result
based on the presence of the dimeric gas is in agreement with the recent
experiment about heat capacity \cite{HEAT}. Finally, we give a brief summary
and discussion in Sev. V.

\section{Mixture of two-component Fermi gases and dimeric gas at zero
temperature}

In the experiment \cite{HEAT} about the heat capacity, the phase transition
for the strongly interacting Fermi gases is observed clearly at a transition
temperature. If there is a phase transition for the emergence of a stable
dimeric gas at a transition temperature, there would be a mixture of
two-component Fermi gases and dimeric gas below the transition temperature,
and the evidence for the mixture of two-component Fermi gases and dimeric
gas would give us an important evidence that there is a phase transition for
the emergence of the stable dimeric gas. To give a clear presentation, we
first propose in this section a theoretical model based on the mixture of
two-component Fermi gases and dimeric gas at zero temperature \cite{XIONGON}%
, while in the following sections we will consider the thermodynamic
properties of this mixture gas at finite temperature.

In the present experiments on two-component Fermi gases, an equal mixture of
Fermi gases is prepared below the Fermi temperature. Assuming that $N$ is
the total number of fermionic atoms without dimers, while $N_{F}$ and $N_{D}$
are the number of fermionic atoms and dimers, we have

\begin{equation}
N=N_{F}+2N_{D}=2N_{F\uparrow }+2N_{D},  \label{number}
\end{equation}%
where $N_{F\uparrow }$ ($=N_{F\downarrow }$) being the number of fermionic
atoms in a spin state, while $N_{F}=2N_{F\uparrow }$ being the total number
of fermionic atoms.

At zero temperature, for the mixture of two-component Fermi gases and
dimeric gas, the dynamic equilibrium is characterized by the fact that the
Gibbs free energy of the system is a minimum. Assuming that $\mu _{F\uparrow
}$ and $\mu _{F\downarrow }$ being the chemical potential of the Fermi gases
and $\mu _{D}$ being the chemical potential of the dimeric gas, the minimum
of the Gibbs free energy means that

\begin{equation}
2\mu _{F\uparrow }=2\mu _{F\downarrow }=\mu _{D}.  \label{chemical-potential}
\end{equation}%
The chemical potential of the dimeric gas is $\mu _{D}=\varepsilon _{D}+\mu
_{t}$ with $\varepsilon _{D}$ being the dimeric binding energy and $\mu _{t}$
being the contribution to the chemical potential due to the thermal
equilibrium of the dimeric gas.

For the unitarity limit that the absolute value $\left\vert a\right\vert $
of the scattering length is much larger than the average distance $\overline{%
l}$\ between atoms, although the scattering length $a$ will play an
important role, it will not appear in the final result of a physical
quantity such as the chemical potential because it can be regarded as
infinity. In this case, the length scale $\overline{l}\sim n_{F\uparrow
}^{-1/3}$ ($n_{F\uparrow }$ is the density distribution of the fermionic
atoms in a spin state) rather than $a$ will appear in the final result of a
physical quantity which means a universal behavior for a system. One can get
a rough expression for the chemical potential through a dimensional analysis
that $\mu _{F\uparrow }\sim \left( \Delta p\right) ^{2}/2m\sim \hbar ^{2}/2m%
\overline{l}^{2}\sim \hbar ^{2}n_{F\uparrow }^{2/3}/2m$. Based on this
dimensional analysis and local density approximation, at zero temperature,
one can get the following form of the chemical potential $\mu _{F\uparrow }$
($=\mu _{F\downarrow }$) for the Fermi gas:

\begin{equation}
\mu _{F\uparrow }=\left( 1+\beta _{F}\right) \frac{\hbar ^{2}\left( 6\pi
^{2}\right) ^{2/3}}{2m}n_{F\uparrow }^{2/3}+V_{ext}\left( \mathbf{r}\right) ,
\label{chemical-fermi}
\end{equation}%
where $V_{ext}\left( \mathbf{r}\right) =m\left( \omega _{x}^{2}x^{2}+\omega
_{y}^{2}y^{2}+\omega _{z}^{2}z^{2}\right) /2$ is the external potential of
the fermionic atoms. Without $\beta _{F}$, the above expression gives the
chemical potential of an ideal Fermi gas in the local density approximation.
The parameter $\beta _{F}$ shows the role of the extremely large scattering
length $a$ in the unitarity limit. The parameter $\beta _{F}$ was first
measured in \cite{THOMAS-Fermi} with a careful experimental investigation of
the strongly interacting two-component Fermi gases near the Feshbach
resonance. $\beta _{F}$ has been also calculated with different theoretical
methods \cite{CARLSON,HEISELBERG}. In \cite{CARLSON}, $\beta _{1}=-0.56$
based on a quantum Monte Carlo calculation. From Eq. (\ref{chemical-fermi}),
the chemical potential at zero temperature is given by%
\begin{equation}
\mu _{F\uparrow }\left( T=0\right) =\mu _{F\downarrow }\left( T=0\right) =%
\sqrt{1+\beta _{F}}T_{F}k_{B},  \label{fermi-chem}
\end{equation}%
where $T_{F}=\left( 6N_{F\uparrow }\right) ^{1/3}\hbar \omega _{ho}/k_{B}$
with $\omega _{ho}=\left( \omega _{x}\omega _{y}\omega _{z}\right) ^{1/3}$.
In getting the above result, we have used a useful mathematical technique
that with the definition of the effective mass $m_{F}^{\ast }=m/\left(
1+\beta _{F}\right) $ and effective angular frequency $\omega _{x}^{\ast }=%
\sqrt{1+\beta _{F}}\omega _{x}$, $\omega _{y}^{\ast }=\sqrt{1+\beta _{F}}%
\omega _{y}$, $\omega _{z}^{\ast }=\sqrt{1+\beta _{F}}\omega _{z}$ so that $%
\mu _{F\uparrow }$ in Eq. (\ref{chemical-fermi}) becomes the expression of
an ideal Fermi gas about the effective mass $m_{F}^{\ast }$ and effective
angular frequency. This mathematical technique for the definition of the
effective mass and effective angular frequency will be used in the whole
paper.

On resonance, because the scattering length between fermionic atoms with
different spin state is divergent, the scattering length $a_{D}$ between
dimers can be also regarded as divergent. In the unitarity limit, for the
dimeric gas, the scattering length $a_{D}$ should not appear in the final
result of a physical quantity too. We assume here that the dimeric gas shows
an analogous universal behavior with the Fermi gas. Through the dimensional
analysis analogous to the Fermi gas in the unitarity limit, the chemical
potential $\mu _{D}$ of the dimeric gas at zero temperature is given by

\begin{equation}
\mu _{D}=\left( 1+\beta _{D}\right) \frac{\hbar ^{2}\left( 6\pi ^{2}\right)
^{2/3}}{2\times 2m}n_{D}^{2/3}+2V_{ext}\left( \mathbf{r}\right) +\varepsilon
_{D},  \label{chemical-dimer}
\end{equation}%
where $n_{D}$ is the density distribution of the dimeric gas. The parameter $%
\beta _{D}$ in $\mu _{D}$ is also due to the large scattering length $a_{D}$
between dimers and it can be calculated in the unitarity limit. Near the
Feshbach resonance, the dimeric energy $\varepsilon _{D}\sim \hbar
^{2}/ma^{2}$. Thus, on resonance, $\varepsilon _{D}$ can be omitted in the
above expression. From the above equation, one can get the chemical
potential for the dimeric gas at zero temperature:

\begin{equation}
\mu _{D}\left( T=0\right) =\sqrt{1+\beta _{D}}T_{D}k_{B},  \label{dimer-chem}
\end{equation}%
where $T_{D}=\left( 6N_{D}\right) ^{1/3}\hbar \omega _{ho}/k_{B}$ with $%
N_{D} $ being the number of dimers. In getting the above result, we have
used a useful mathematical technique that the chemical potential becomes the
form of an ideal gas about the effective mass $m_{D}^{\ast }=2m/\left(
1+\beta _{D}\right) $ and effective angular frequency $\omega _{x}^{\ast }=%
\sqrt{1+\beta _{F}}\omega _{x}$, $\omega _{y}^{\ast }=\sqrt{1+\beta _{F}}%
\omega _{y}$, $\omega _{z}^{\ast }=\sqrt{1+\beta _{F}}\omega _{z}$ of the
dimer.

From the dynamic equilibrium condition $2\mu _{F\uparrow }=\mu _{D}$ and the
chemical potential given by Eqs. (\ref{fermi-chem}) and (\ref{dimer-chem}),
one gets the following general equation to determine the fraction of dimers
at zero temperature and on Feshbach resonance:

\begin{equation}
4\beta _{r}\left( 1-x\right) ^{2/3}=x^{2/3},  \label{fraction}
\end{equation}%
where $\beta _{r}=\left( 1+\beta _{F}\right) /\left( 1+\beta _{D}\right) $
and the fraction of dimers $x=2N_{D}/N$.

In the presence of dimers, the chemical potential of the system is obviously
lower than the case without dimers. Thus, the mixture gas of the fermionic
atoms and dimers is a stable state of the system. For $\left\vert
a\right\vert >>\overline{l}$, the fraction of the dimeric gas is a general
result once the system is in the unitarity limit. One can see easily from
Eq. (\ref{fraction}) that there is a quite high fraction of dimers if the
value of\textit{\ }$\beta _{D}$ is close to $\beta _{F}$. An indirect
evidence of the dimeric gas is given by a recent experiment \cite%
{MIT-fermion} that there is significant fraction of the molecules in
zero-momentum state after a fast magnetic field transfer (the magnetic field
is swept below the Feshbach resonant magnetic field $B_{0}$ so that the
scattering length between fermionic atoms becomes positive) to create bound
molecules from the dimers. After the dimer-molecule conversion, in \cite%
{MIT-fermion} the maximum fraction of the molecules in zero-momentum state
is observed to be $80\%$. In this case, $\beta _{r}$ is estimated to be $%
0.63 $.

At zero temperature and thus the thermal excitation is omitted, the dimer
comprising two fermionic atoms is quite stable. Because the dimeric gas is
immersed in the degenerate two-component Fermi gases, due to Pauli blocking
comes from the degenerate Fermi gases, any dimer can not be dissociated into
two fermionic atoms once the equilibrium is attained so that $2\mu
_{F\uparrow }=\mu _{D}$. In fact, the stability of the dimeric gas is
consistent with the experiment \cite{MIT-fermion} that extremely close to
the Feshbach resonance there is no obvious decreasing of the molecules in
zero-momentum state after the dimer-molecule conversion process even after $%
10$ \textrm{s }hold time of the final magnetic field. Due to the stability
and Pauli blocking, the dimer-molecule conversion will always convert the
fermionic atom pairs in a same dimer into bound molecules. Thus at zero
temperature, all the dimers will be converted into molecules with
zero-momentum state during the dimer-molecule conversion process. This means
that the fraction of dimers in thermal equilibrium investigated here can be
used to explain the experimental result of the fraction of zero-momentum
molecules after the dimer-molecule conversion.

In the recent experiments \cite{THOMAS,GRIMM1}, the frequency of a radial
breathing mode is observed extremely close to the Feshbach resonance and
found to agree well with the theoretical predication based on hydrodynamic
theory in the unitarity limit \cite{STRI}. Combining with the experimental
result in \cite{MIT-fermion} that there is a high fraction of molecules in
zero-momentum state after the dimer-molecule conversion process, we see that
the dimeric gas should play a dominant role in determining the frequency of
the collective oscillations due to its high fraction. For the radial
breathing mode, the agreement of the experiment with theoretical result in
the unitarity limit shows that it is reasonable to describe the dimeric gas
by the chemical potential given by Eq. (\ref{chemical-dimer}). In fact, an
analogous form of Eq. (\ref{chemical-dimer}) is used to calculate the
frequency of the radial breathing mode in \cite{STRI}, whose result agrees
well with the experiment in \cite{THOMAS,GRIMM1}. Different from the
theoretical model in \cite{STRI}, however, in the present work on resonance,
the dimeric gas plays a dominant role in the frequency of the collective
oscillations at zero temperature, rather than the Fermi gases.

\section{Phase transition for the emergence of stable dimeric gas}

On resonance, the divergent scattering length between dimers would make the
property of the dimeric gas become quite unique. From the form of the
chemical potential (\ref{chemical-dimer})\textit{\ }for the dimeric gas
which is consistent with the experiments about the condensate fraction \cite%
{MIT-fermion} and collective excitations \cite{THOMAS,GRIMM1}, the dimeric
gas is described by a density of states analogous to that of a Fermi gas.
This unique property of the dimeric gas is due to the divergent scattering
length between dimers, and is enlightened by the well-known result that a
one-dimensional strongly interacting Bose gas behave like Fermi gases \cite%
{GIRARDEAU,LIEB} which has been verified by recent experiments \cite%
{BLOCH,WEISS}. Based on these considerations, we assume here that the dimers
with divergent scattering length would be described by the Fermi statistics
although the dimer is a boson. Obviously, the validity of this assumption
should be finally tested by experiments.

Based on the above assumption that the dimers satisfy Fermi statistics, at
finite temperature, the occupation number of the dimers at a state of the
energy level $\varepsilon _{D}$ is given by

\begin{equation}
f_{D}\left( \varepsilon _{D}\right) =\frac{1}{e^{\left( \varepsilon _{D}-\mu
_{D}\left( T\right) \right) /k_{B}T}+1},  \label{f-dimer}
\end{equation}%
while the occupation number of the fermionic atoms at a state of the energy
level $\varepsilon _{F}$ is given by

\begin{equation}
f_{F\uparrow }\left( \varepsilon _{F}\right) =f_{F\downarrow }\left(
\varepsilon _{F}\right) =\frac{1}{e^{\left( \varepsilon _{F}-\mu _{F\uparrow
}\left( T\right) \right) /k_{B}T}+1}.  \label{f-fermi}
\end{equation}%
In Eqs. (\ref{f-dimer}) and (\ref{f-fermi}), the strong interaction has been
considered through the chemical potential $\mu _{D}$ and $\mu _{F\uparrow }$
which depends respectively on the parameters $\beta _{D}$ and $\beta _{F}$.

For the state below the energy level $\varepsilon _{F}$ for the fermionic
atoms and $\varepsilon _{D}$ for the dimers, if the occupation number of the
fermionic atoms and dimers in a state are both $1$, the dimers can not be
dissociated into two fermionic atoms below the energy level $\varepsilon
_{F} $ because of Pauli exclusion principle. However, it is possible that
the dimers can be thermally excited to high energy level and dissociated
into two fermionic atoms especially because the binding energy of the dimers
can be omitted on resonance. Thus, sufficient low temperature is needed for
the existence of a stable dimeric gas. Assuming that $\varepsilon _{F}^{c}$
and $\varepsilon _{D}^{c}$ are the critical energy level that below these
energy level there is a stable coexistence of two-component Fermi gases and
dimeric gas, due to the fact that the dimer comprises two fermionic atoms
with different spin state, one has $\varepsilon _{D}^{c}=2\varepsilon
_{F}^{c}$. One can also understand this relation from the chemical
equilibrium condition $\mu _{D}=2\mu _{F\uparrow }$. The stability of the
coexistence of these gases in the presence of thermal excitation would give
us a strong confinement condition on the value of $\varepsilon _{F}^{c}$ and
$\varepsilon _{D}^{c}$. At finite temperature, the thermally excited energy
for fermionic atom and dimer is respectively given by $\varepsilon
_{therm}^{F}=3k_{B}T/2$ and $\varepsilon _{therm}^{D}=3k_{B}T$. $\varepsilon
_{therm}^{D}/\varepsilon _{therm}^{F}=2$ is due to the fact that a dimer
comprises two fermionic atoms.

Based on the above analyses, $\varepsilon _{F}^{c}$ and $\varepsilon
_{D}^{c} $ can be determined through the following equation:\bigskip
\begin{equation}
2f_{F\uparrow }\left( \varepsilon _{F}^{c}+\varepsilon _{therm}^{F}\right)
+2f_{D}\left( \varepsilon _{D}^{c}+\varepsilon _{therm}^{D}\right) =3.
\label{transition}
\end{equation}%
The left hand side of the above equation shows the overall number of atoms
(Note that a dimer comprises two fermionic atoms). The factor $3$ on the
right hand side of the above equation is due to Pauli exclusion principle
for the fermionic atoms and the assumption that the dimer satisfies Fermi
statistics in this paper. When the confinement condition given by Eq. (\ref%
{transition}) is satisfied, the dimers below the energy level $\varepsilon
_{D}^{c}$ can not be effectively thermally excited and dissociated into two
fermionic atoms.

From Eqs. (\ref{f-dimer}) and (\ref{f-fermi}), Eq. (\ref{transition}) can be
given more explicitly by

$${
\frac{2}{e^{\left[ \varepsilon _{F}^{c}+\frac{3}{2}k_{B}T-\mu
_{F\uparrow }\left( T\right) \right] /k_{B}T}+1}+ }$$

\begin{equation}
\frac{2}{e^{\left[ \varepsilon _{D}^{c}+3k_{B}T-\mu _{D}\left(
T\right) \right] /k_{B}T}+1}=3. \label{explict}
\end{equation}%
At finite temperature, the chemical potential $\mu _{F\uparrow }\left(
T\right) $ is estimated to be:

\begin{equation}
\mu _{F\uparrow }\left( T\right) =\mu _{F\uparrow }\left( T=0\right) \left[
1-\frac{\pi ^{2}}{3}\left( \frac{k_{B}T}{\mu _{F\uparrow }\left( T=0\right) }%
\right) ^{2}\right] .  \label{expansion}
\end{equation}%
The transition temperature for the emergence of a stable dimeric gas is then
obtained by setting $\varepsilon _{F}^{c}=0$ and $\varepsilon _{D}^{c}=0$ in
Eq. (\ref{explict}). Using $\mu _{D}\left( T\right) =2\mu _{F\uparrow
}\left( T\right) $ and assuming that $\xi =k_{B}T/\mu _{F\uparrow }\left(
T=0\right) =T/\sqrt{1+\beta _{F}}T_{F}$, the critical value $\xi _{0}$\ is
then determined by

\begin{equation}
\frac{2}{e^{\left[ 3/2-\left( 1-\pi ^{2}\xi _{0}^{2}/3\right) /\xi _{0}%
\right] }+1}+\frac{2}{e^{\left[ 3-2\left( 1-\pi ^{2}\xi _{0}^{2}/3\right)
/\xi _{0}\right] }+1}=3.
\end{equation}%
In this case, one find that $\xi _{0}=0.31$, which agrees very well with the
experimental result $0.33$\ \cite{HEAT}. This theoretical value of $\xi _{0}$
also shows that one can use the approximate expression for the chemical
potential $\mu _{F\uparrow }\left( T\right) $ at finite temperature given by
Eq. (\ref{expansion}) because for this value of $\xi _{0}$ the high-order
term of the chemical potential $\mu _{F\uparrow }\left( T\right) $ can be
omitted.

Below the transition temperature, one can get the critical energy level for
the dimers from Eq. (\ref{explict}) which is given by

$${
\varepsilon _{D}^{c}=\mu _{D}\left( T=0\right) \left[ 1+\left(
\frac{\pi ^{2}\xi _{0}}{3}-\frac{1}{\xi _{0}}\right)
\frac{2k_{B}T}{\mu _{D}\left( T=0\right) }\right. }$$

\begin{equation}
\left.-\frac{4\pi ^{2}}{3}\left( \frac{k_{B}T}{\mu _{D}\left(
T=0\right) }\right) ^{2}\right] .
\end{equation}%
When the dimers satisfy Fermi statistics, the density state of the dimers is
proportional to $\varepsilon _{D}^{2}$. The relation between the number of
stable dimers and temperature is then given by

\begin{equation}
N_{D}\left( T\right) =N_{D}\left( T=0\right) \left[ 1+\left( \frac{\pi
^{2}\xi _{0}}{3}-\frac{1}{\xi _{0}}\right) \xi -\frac{\pi ^{2}\xi ^{2}}{3}%
\right] ^{3},  \label{dimer-number}
\end{equation}%
where $N_{D}\left( T=0\right) $\ is the number of dimers at zero temperature.

\section{The overall entropy of two-component Fermi gases and dimeric gas}

\bigskip The entropy of two-component Fermi gases takes the following form%
\begin{equation}
S_{F}=-2k_{B}\sum_{\lambda _{F\uparrow }}\left[ f_{\lambda _{F\uparrow }}\ln
f_{\lambda _{F\uparrow }}+\left( 1-f_{\lambda _{F\uparrow }}\right) \ln
\left( 1-f_{\lambda _{F\uparrow }}\right) \right] ,  \label{entropy-fermi}
\end{equation}%
where $\lambda _{F\uparrow }$ denotes the state of the fermionic atom and $%
f_{\lambda _{F\uparrow }}$ is the occupation number of the fermionic atom in
the state $\lambda _{F\uparrow }$. The factor $2$ is due to two spin states
of the fermionic atom. For the dimeric gas, under the assumption that the
dimers satisfy the Fermi statistics, the entropy of the dimeric gas is given
by

\begin{equation}
S_{D}=-k_{B}\sum_{\lambda _{D}}\left[ f_{\lambda _{D}}\ln f_{\lambda
_{D}}+\left( 1-f_{\lambda _{D}}\right) \ln \left( 1-f_{\lambda _{D}}\right) %
\right] ,  \label{entropy-dimer}
\end{equation}%
where $\lambda _{D}$ denotes the state of the dimers and $f_{\lambda _{D}}$
is the occupation number of the dimer in the state $\lambda _{D}$. The
overall entropy $S_{F-D}$ of the system is then given by

\begin{equation}
S_{F-D}=S_{F}+S_{D}.  \label{overall-entropy}
\end{equation}

To show the role of the dimeric gas in the thermodynamic properties of the
system, we first consider the entropy of the system for the temperature far
below the Fermi temperature $T_{F}$ (or $T_{D}$ for the dimeric gas). In
this case, to the first order approximation, the overall entropy $S_{F-D}$
is given by

\begin{equation}
S_{F-D}=\frac{\pi ^{2}N_{F}k_{B}T}{\sqrt{1+\beta _{F}}T_{F}}+\frac{\pi
^{2}N_{D}k_{B}T}{\sqrt{1+\beta _{D}}T_{D}}.
\end{equation}%
In the above equation for the entropy, the second term represents the
contribution of the dimeric gas. Using the condition of the chemical
equilibrium given by Eq. (\ref{chemical-potential}), one has

\begin{equation}
S_{F-D}=\frac{\pi ^{2}\left( 2N_{F}+N_{D}\right) k_{B}T}{2\sqrt{1+\beta _{F}}%
T_{F}}.
\end{equation}

Using the formula $\partial S_{F-D}/\partial E=1/T$, we have

\begin{equation}
\frac{\partial S_{F-D}}{\partial E}=\frac{\partial S_{F-D}}{\partial T}\frac{%
\partial T}{\partial E}=\frac{1}{T}.
\end{equation}%
Thus

\begin{equation}
\frac{\partial E}{\partial T}=T\frac{\partial S_{F-D}}{\partial T}.
\label{energy-partial}
\end{equation}%
In the limit of zero temperature, omitting the high order terms of the
partial differential $\partial S_{F-D}/\partial T$, one has

\begin{equation}
\frac{\partial S_{F-D}}{\partial T}=\frac{\pi ^{2}\left( 2N_{F}+N_{D}\right)
k_{B}}{2\sqrt{1+\beta _{F}}T_{F}}.
\end{equation}

From the above equation, we get

\begin{equation}
\frac{\partial E}{\partial T}=\frac{\pi ^{2}\left( 2N_{F}+N_{D}\right) k_{B}T%
}{2\sqrt{1+\beta _{F}}T_{F}}.
\end{equation}%
In the limit of zero temperature, $E$ is then given by:

\begin{equation}
E=E_{0}+\frac{\pi ^{2}\left( 2N_{F}+N_{D}\right) k_{B}T^{2}}{4\sqrt{1+\beta
_{F}}T_{F}},  \label{totoal}
\end{equation}%
where $E_{0}$ is the total ground state energy of the mixture gases at zero
temperature. After a simple calculation, one gets

\begin{equation}
E_{0}=\frac{3N\sqrt{1+\beta _{F}}k_{B}T_{F}}{4}.  \label{E0}
\end{equation}

From Eqs. (\ref{totoal}) and (\ref{E0}), we have

\begin{equation}
\frac{E}{E_{0}}-1=\frac{2\pi ^{2}\alpha }{3}\left( \frac{T}{T_{F}}\right)
_{fit}^{2},  \label{factor}
\end{equation}%
where $\alpha =1-3N_{D}\left( T\right) /2N$ and $\left( T/T_{F}\right)
_{fit}\equiv \xi =T/\sqrt{1+\beta _{F}}T_{F}$. Below the transition
temperature for the emergence of stable dimeric gas, from Eq. (\ref%
{dimer-number}), one has

\begin{equation}
\alpha \left( T\right) =1-\frac{3N_{D}\left( T=0\right) }{2N}\left[ 1+\left(
\frac{\pi ^{2}\xi _{0}}{3}-\frac{1}{\xi _{0}}\right) \xi -\frac{\pi ^{2}\xi
^{2}}{3}\right] ^{3}.  \label{T-alpha}
\end{equation}

To compare with the experimental result in \cite{HEAT}, from Eq. (\ref%
{factor}), one gets\bigskip
\begin{equation}
\ln \left( \frac{E}{E_{0}}-1\right) =\ln \left[ \frac{2\pi ^{2}}{3}\left(
\frac{T}{T_{F}}\right) _{fit}^{2}\right] +\ln \alpha .  \label{lnalpha}
\end{equation}%
\bigskip Without the factor $\ln \alpha $, the above equation gives the
result for strongly interacting two-component Fermi gases without the
dimeric gas. Thus, the factor $\ln \alpha $ on the right hand side of the
above equation will give us clearly the role of the dimeric gas in the
thermodynamic properties of the system. Below the transition temperature
(i.e. in the presence of the dimeric gas), $\alpha $ is smaller than $1$ and
thus $\ln \alpha $ is negative. In this case, compared with the theoretical
model without dimeric gas, in the relation between $\ln \left(
E/E_{0}-1\right) $ and $\ln \left( T/T_{F}\right) _{fit}$, the presence of
the dimeric gas will decrease significantly the value of $\ln \left(
E/E_{0}-1\right) $. In the experimental measurement of the condensate
fraction on resonance in \cite{MIT-fermion}, $2N_{D}\left( T=0\right) /N$
can be estimated to be $0.8$ in the limit of zero temperature. In this case,
the factor $\alpha $ is estimated to be $0.4$. Thus in the limit of zero
temperature, one has $\ln \alpha =-0.92$ which means a significant
decreasing of $\ln \left( E/E_{0}-1\right) $ when the presence of dimeric
gas is considered. In fact, this effect has been observed in the experiment
\cite{HEAT} which shows clearly the phase transition for the emergence of
the dimeric gas through the investigation of the relation between $\ln
\left( E/E_{0}-1\right) $ and $\ln \left( T/T_{F}\right) _{fit}$.

\begin{figure}[tbp]
\includegraphics[width=0.75\linewidth,angle=270]{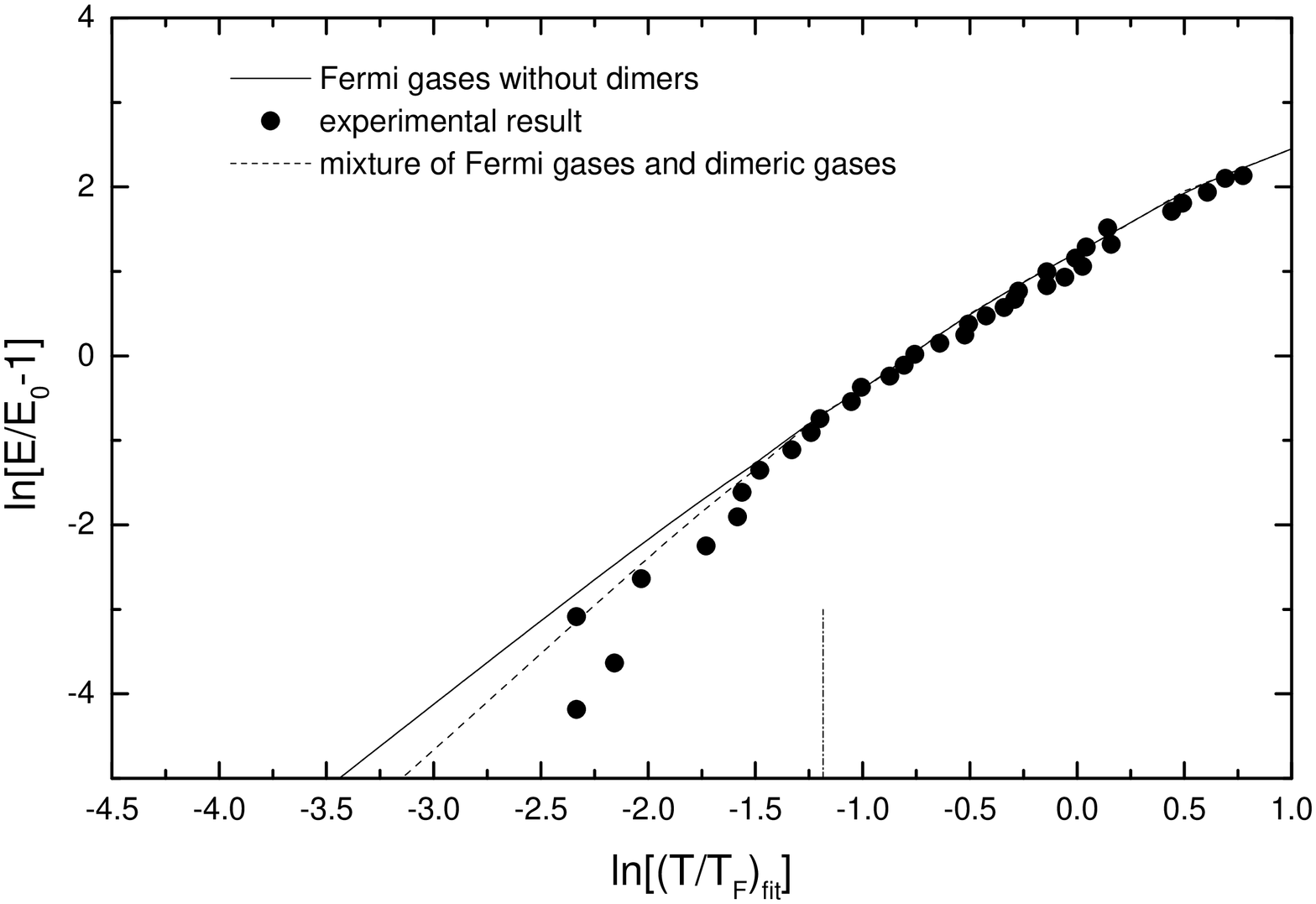}
\caption{Shown in figure 1 is the theoretical and experimental
results of the relation between $\ln \left( E/E_{0}-1\right) $ and
$\ln \left( T/T_{F}\right) _{fit}$. The circles show the
experimental result by J. Kinast \textit{et al }\cite{HEAT}, while
the solid line is the theoretical predication of the strongly
interacting two-component Fermi gases without the consideration of
the dimeric gas. It is clearly shown that there is a significant
difference between the solid line and experimental data at low
temperature which means that there is a phase transition for the
strongly interacting Fermi gases. The dashed line is the
theoretical result based on the consideration that there is a
mixture of the Fermi gases and dimeric gas below a transition
temperature. We see that this theoretical model agree well with
the experimental result. The dash-dotted line shows the transition
temperature $T_{tran}=0.31\sqrt{1+\beta _{F}}T_{F}$ based on the
consideration of the thermal excitation of the dimers, which is in
agreement with the experimental result $0.33\sqrt{1+\beta
_{F}}T_{F}$.}
\end{figure}

We now give a comparison of our theoretical result with the recent
experimental result in \cite{HEAT}. In this experiment by Kinast \textit{et
al}, an equal mixture of the degenerate Fermi gases is prepared for two
lowest spin states of $^{6}Li$ atoms, and the magnetic field is tuned to be $%
840$ \textrm{G} which is just above the Feshbach resonant magnetic field.
For the magnetic field of $840$ \textrm{G}, the absolute value of the
scattering length is much larger than the average distance between particles
so that our theoretical model can be used to analysis the experimental
result. From Eqs. (\ref{overall-entropy}), (\ref{energy-partial}) and (\ref%
{E0}), the dashed line shown in Fig.1 is the numerical result of the
relation between $\ln \left( E/E_{0}-1\right) $ and $\ln \left(
T/T_{F}\right) _{fit}$ by using the transition temperature $T_{tran}=0.31%
\sqrt{1+\beta _{F}}T_{F}$ obtained in this paper. The circles show the
experimental result in \cite{HEAT}, while the solid line shows the
theoretical result based on the model of two-component Fermi gases without
dimeric gas. We see that below the transition temperature, the theoretical
model based the mixture of two-component Fermi gases and dimeric gas agrees
with the experimental result.

\section{Summary and Discussion}

In summary, we investigate the thermodynamic properties of the strongly
interacting two-component Fermi gases by considering specially the role of
the dimeric gas. Due to the divergent scattering length between dimers, we
assume that the dimers satisfy Fermi statistics and there is a university
behavior for the dimeric gas. It is found that this simple theoretical model
agrees with the recent experiments about the condensate fraction, collective
excitations, transition temperature and energy of the ultracold gases near
the magnetic-field Feshbach resonance. The model proposed here is a
phenomenological theory just like the effective force is introduced to
describe the atomic nucleus which is a complex, many-body system bound by
the strong interaction. Nevertheless, the agreement with the experimental
results gives us a strong support for the validity of our theoretical model
although the microscopic origin of our theory is still remained to be
revealed. In future work, we will try to find the microscopic mechanism of
the theoretical model proposed here.

Note added.$-$In preparing this manuscript, we noticed a recent work by Q.
Chen \textit{et al} \cite{QCHEN} where a theoretical model is proposed based
on the mixture of fermionic atoms and hybridized bosons. There is an
astonishing consistency between the theoretical result obtained by Q. Chen
\textit{et al} \cite{QCHEN} and our theoretical result although the dimer is
assumed by us to satisfy Fermi statistics due to its divergent scattering
length.

\begin{acknowledgments}
We thank Prof. J. E. Thomas for useful discussions. This work is
supported by the National Natural Science Foundation of China
under Grant Nos. 10205011, 10474117, 10474119, the National Basic
Research Programme of China under Grant No.001CB309309, and also
by funds from the Chinese Academy of Sciences.
\end{acknowledgments}




\end{document}